\documentclass[fleqn,usenatbib]{mnras}

\usepackage{graphicx,color}
\usepackage{amsmath}
\usepackage{amssymb}

\newcommand\msun{{\,{\rm M}_{\odot}}}
\newcommand\simlt{\la}

\newcommand\be{\begin{equation}}
\newcommand\ee{\end{equation}}

\title[Fragmentation of turbulent discs]{On fragmentation of turbulent self-gravitating discs in the long cooling time regime} 
\author[Ken Rice \& Sergei Nayakshin]{Ken Rice$^{1,2}$\thanks{E-mail: wkmr@roe.ac.uk}, Sergei Nayakshin$^{3}$\\
$^{1}$Scottish Universities Physics Alliance (SUPA), Institute for Astronomy, University of Edinburgh, Blackford Hill, Edinburgh, EH9 3HJ \\
$^{2}$Centre for Exoplanet Science,  University of Edinburgh, Edinburgh, UK \\
$^{3}$Department of Physics and Astronomy, University of Leicester, Leicester, LE17RH, UK}

\date{Accepted XXX. Received YYY; in original form ZZZ}

\pubyear{2017}

\begin{document}
\label{firstpage}
\pagerange{\pageref{firstpage}--\pageref{lastpage}}
\maketitle

\begin{abstract}
It has recently been suggested that in the presence of driven turbulence discs may be much less stable against gravitational collapse than their non turbulent analogs, due to stochastic density fluctuations in turbulent flows. This mode of fragmentation would be especially important for gas giant planet formation. Here we argue, however, that stochastic density fluctuations due to turbulence do not enhance gravitational instability and disc fragmentation in the long cooling time limit appropriate for planet forming discs. These fluctuations evolve adiabatically and dissipate away by decompression faster than they could collapse. We investigate these issues numerically in 2D via shearing box simulations with driven turbulence and also in 3D with a model of instantaneously applied turbulent velocity kicks. In the former setting turbulent driving leads to additional disc heating that tends to make discs more, rather than less, stable to gravitational instability. In the latter setting, the formation of high density regions due to convergent velocity kicks is found to be quickly followed by decompression, as expected. We therefore conclude that driven turbulence does not promote disc fragmentation in protoplanetary discs and instead tends to make the discs more stable. We also argue that sustaining supersonic turbulence is very difficult in discs that cool slowly. 
\end{abstract}


\section{Introduction} \label{Introduction}

Fragmentation of self-gravitating gaseous discs into self-bound clumps is a physically attractive model for the formation of planets, brown dwarfs and low-mass secondary stars orbiting their primary stars \citep[e.g.,][]{kuiper51}. However, detailed models of the process show that gas discs need to be massive, cold, and also need to cool rapidly \citep[e.g.,][]{gammie01,rice05,rafikov05} in order to fragment. For the same reasons, analytic models of the disc tend to give minimum fragment masses close to the brown dwarf regime \citep{kratter10,forgan11}, although there is a significant uncertainty in these estimates  \citep[e.g., see Fig. 3 in][]{kratter16}. On balance, the formation of gas giant planets via gravitational disc instability, as opposed to brown dwarfs or secondary stars, remains controversial.

\cite{hopkins13}, however, recently pointed out that the studies quoted above assumed that the only turbulence was the gravito-turbulence due to the gravitational instability itself. In contrast, in a medium with other sources of turbulence, such as MRI \citep{balbus91}, there are local stochastic density fluctuations which, although rare, could be very significant and could be dense enough to occasionally yield gravitational collapse. The authors developed an analytical theory of turbulent discs which suggested that discs with supersonic turbulence can fragment ``always" -- e.g., at Toomre parameters much larger than unity,  provided that the disc {\it lifetimes are} sufficiently long. This result is also potentially significant because the mass spectrum of the objects formed by the instability is not narrowly peaked at the local Jeans mass but is instead very broadly extended around it, reaching masses up to two orders of magnitude smaller than the Jeans mass. They showed that such a mode of fragmentation could form gaseous bound planets with masses as small as a few Earth masses in the inner few AU of protoplanetary discs.

In this paper we question a key  assumption behind the model of \cite{hopkins13}. Their scenario assumes similarity with turbulence in star forming regions, where gas cools rapidly and can usually be assumed to be isothermal \citep[e.g.,][]{ostriker99,padoan02,federrath10}. However, realistic protoplanetary discs have long cooling times in the regions inside $R \simlt 50-100$~AU \citep{rafikov05,clarke09,kratter16}. Measured in terms of the local dynamical time, $1/\Omega$ (where $\Omega = \sqrt{GM_*/R^3}$ is the local Keplerian angular frequency) the cooling time $\tau_{\rm c} \Omega = \beta \gg 1$, where $\beta$ is a dimensionless number. In this context, we take $\beta \gg 1$ to be $\beta \sim 10$, or greater. This implies that compression generated by small scale  (meaning length scales of the order of the disc height scale, $H$) super-sonic turbulence will evolve adiabatically rather than isothermally.

Here we use two different numerical methods, one with  grid based hydrodynamics (\S \ref{Method}), and one based on Smoothed Particle Hydodynamics (SPH), discussed in Section \ref{sec:velkick}, to test these ideas numerically. We also use two physically different settings to try and shed more light on this problem. In the former we investigate a continuously driven turbulence model, whereas in the latter we impose instantaneous velocity kicks. In both cases gas cooling is modeled via the already mentioned fixed $\beta$-parameter cooling prescription.

\section{Method} \label{Method}

\subsection{Numerical code}

To investigate the evolution of self-gravitating accretion discs in the presence of continuously driven turbulence, we use the fixed grid {\small PENCIL CODE}. The {\small PENCIL CODE} is a finite
difference code that uses sixth-order centred spatial derivatives and a third-order Runge-Kutta time-stepping scheme (see \citealt{brandenburg03} for details).
We use the standard `shearing sheet' approximation (e.g., \citealt{gammie01,rice11}) in which the disc dynamics is studied in a local Cartesian frame
co-rotating with the same angular velocity, $\Omega$, of the disc at some radius from the central star. We assume that the disc is undergoing Keplerian
rotation and so assume a shear parameter of $q = 1.5$.  This means that the $y$-component of the fluid velocity is $u_y = -q \Omega x$.  We also
assume that the unperturbed background surface density, $\Sigma_o$, and the unperturbed two-dimensional pressure, $P_o$, are spatially constant,
and we include a Coriolis force to include the effects of the coordinate frame rotation.

Although we do carry out some isothermal simulations, we have mainly focused on simulations in which the gas has an adiabatic equation of state, 
\begin{equation}
P = (\gamma - 1) U,
\label{eq:eos}
\end{equation}
where $P$ is the two-dimensional pressure, $U$ is the two-dimensional internal energy per unit volume, and $\gamma$ is the two-dimensional adiabatic index, which we take to be $\gamma = 2$ \citep{gammie01}.  

For simulations with an energy equation, we assume that the system cools with a cooling time,
$\tau_c$, that is taken to be constant \citep{gammie01}. The {\small PENCIL CODE} actually solves for the specific entropy, $s$, and so the cooling term in
the energy equation becomes
\begin{equation}
\Sigma T \frac{\partial s}{\partial t} = - \frac{\Sigma c_s^2}{\gamma (\gamma - 1) \tau_c},
\label{eq:energeq}
\end{equation}
where $c_s$ is the local sound speed. We can write the cooling time as $\tau_c = \beta \Omega^{-1}$, where is $\beta$ is again a constant.

There are indications that some simulations that attempt to quantify the fragmentation boundary (i.e., the cooling time below which 
a disc will fragment, rather than sustain a quasi-steady self-gravitating state) are not fully converged \citep{meru11,meru12}. However, it seems that this may primarly be a consequence of the numerical method, rather than an indication that fragmentation can actually occur at much longer cooling times 
than initially suggested \citep{gammie01}.  

For example, it could be a consequence of the form of the cooling implementation \citep{rice12}, or may be related to the artificial viscosity \citep{lodato11,rice14}.  In particular, \citet{deng17} suggest that the artificial viscosity in
Smoothed Particle Hydrodynamics (SPH) can act to artificially remove angular momentum from dense regions, promoting fragmentation. 
Recent numerical simulations \citep{baehr17} have also demonstrated convergence, and the suggested fragmentation boundary is also consistent with earlier results \citep{gammie01,rice05} and with
semi-analytic calculations \citep{lin2016}. It is, therefore, largely accepted that disc fragmentation requires $Q \sim 1$ and cooling times that are comparable to,
or shorter than, the local dynamical timescale. 

For $\gamma = 2$ we would expect that a self-gravitating accretion disc would be unable to sustain a state of marginal stability \citep{paczynski78}
for cooling times in which $\beta \le 3$ \citep{gammie01, rice05}. In such a case, we would expect the disc to fragment to form a number of bound
objects, potentially planetary-mass bodies in discs around young stars.

\subsection{Initial conditions} \label{init_conds}

All the {\small PENCIL CODE} simulations below are performed in a rectangular box of size $L_x = L_y = 320$ with a resolution $N \times N$, where $N = 1024$. We work in the system of units in which $G=1$ and $\Omega =1$.
The sound speed, $c_s$, and surface density, $\Sigma$, are initially constant and are set so that $Q_{\rm init}  = c_s \Omega/\pi G \Sigma = 1.5$.
We take the initial surface density to be $\Sigma = 1$, so that the sound speed is initially $c_s = 1.5 \pi$.
With  these initial conditions, the box is large enough to properly represent the spiral density waves, which will have length scales much larger than the disc vertical scale height, $H = c_s/\Omega = 1.5 \pi$, while also resolving the Jeans length when $Q \sim 1$. 

The simulations are seeded with initial perturbations introduced through the velocity field, which is perturbed from the
background flow via a Gaussian noise component with a subsonic amplitude. All the simulations are run for a total of 500 time units. Note that with the velocity shear imposed, any structure at the edge of the box (i.e., $|x| \sim 160$) will cross the box in less than 2 time units.

For simulations that use the energy equation, we also prescribe a constant cooling
time $\tau_c$, which is equivalent to $\beta$, given that $\Omega = 1$.  As discussed below, the turbulence has a prescribed amplitude and forcing
wavenumber.  Our simulations are initially evolved for 50 time units without any cooling. This is simply to ensure that it has some time to settle before imposing any cooling. The cooling is then imposed, with initially a large $\tau_c$ which then decays to the prescribed value of $\tau_c$ for a given simulation over the next 50 time units. In these simulations the instability often undergoes an initial transient burst phase before settling towards a quasi-steady state, and this relaxation procedure avoids the system artificially fragmenting during the initial transient phase. The simulation is then run for a further 100 time units before imposing the driven turbulence (i.e.,
the turbulent forcing is typically turned on at $t = 200$).  This, again, ensures that the system is well past the transient burst phase and is likely close to the final state to which it would settle in the absence of an additional turbulent forcing, before introducing the additional turbulent forcing. We will specify any simulations below that deviate from this initialization procedure.

\subsection{Turbulent forcing} \label{turb_forcing}

We wish to use a simple mathematical model to describe turbulent disc flows. For the {\small PENCIL CODE} simulations we use a turbulent forcing method first introduced by \citet{haugen04}. In this method, a forcing function, $\textbf{\textit{f}}$, which depends on position and time, describes the local acceleration that gas experiences due to turbulent eddy motions. In this approach, the forcing function has the form

\begin{equation}
\textbf{\textit{f}}(\textbf{\textit{x}},t) = {\mathrm Re} \left\{ N \textbf{\textit{f}}_{\textbf{\textit{k}}(t)} \exp \left[{\mathrm i} \textbf{\textit{k}}(t) \cdot \textbf{\textit{x}} + {\mathrm i} \phi (t) \right] \right\},
\label{eq:forcing}
\end{equation}
where $\textbf{\textit{x}}$ is the position vector.  The wave vector, $\textbf{\textit{k}}(t)$, and the random phase, $ -\pi < \phi(t) < \pi$, change at each time step. 
So that the time integrated forcing function is independent of the time step $\delta t$, the normalisation factor, $N$, has to be proportional to $\delta t^{1/2}$. 
As in \citet{haugen04}, we take it to be $N = f_o c_s \left( | \textbf{\textit{k}} | c_s/\delta t \right) ^{1/2}$.  To vary the strength of the forcing, we vary 
the coefficient $f_o$. 

In each simulation we specify the magnitude of the forcing wavenumber, $| \textbf{\textit{k}} |$, but at each timestep we randomly select the direction of this 
wavevector. We then force the system with nonhelical, transverse waves
\begin{equation}  
\textbf{\textit{f}}_\textbf{\textit{k}} = \frac{\left( \textbf{\textit{k}} \times \textbf{\textit{e}} \right) }{\sqrt{\textbf{\textit{k}}^2 - \left( \textbf{\textit{k}} \cdot \textbf{\textit{e}} \right)^2}},
\label{eq:waves}
\end{equation}
where $\textbf{\textit{e}}$ is an arbitrary vector not aligned with $\textbf{\textit{k}}$.  Given that our simulations are two-dimensional, and we want
these to be transverse waves, $\textbf{\textit{e}}$ is taken to be a unit vector in the $z$-direction.

\begin{figure}
\begin{center}
\includegraphics[scale=0.45]{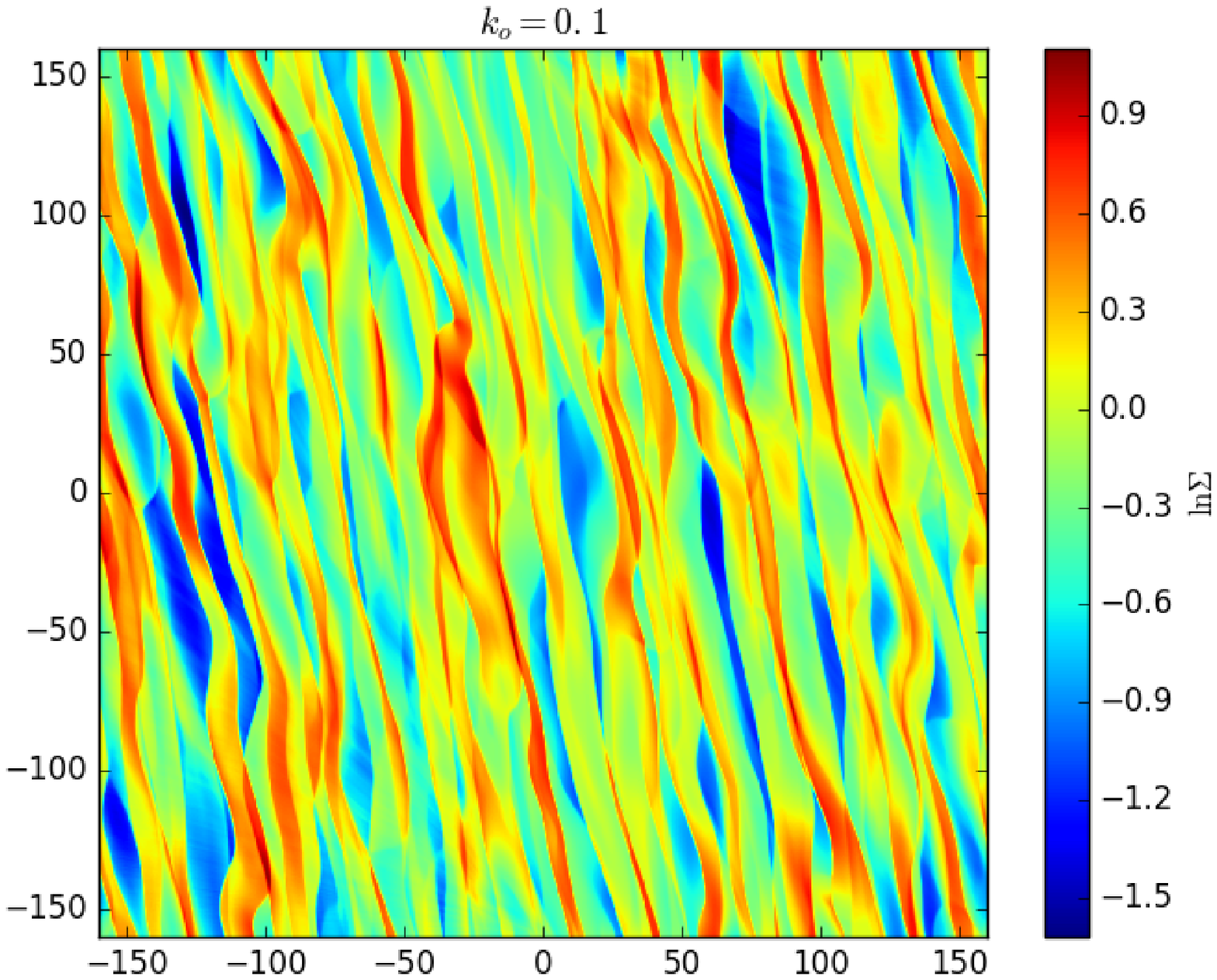}
\includegraphics[scale=0.45]{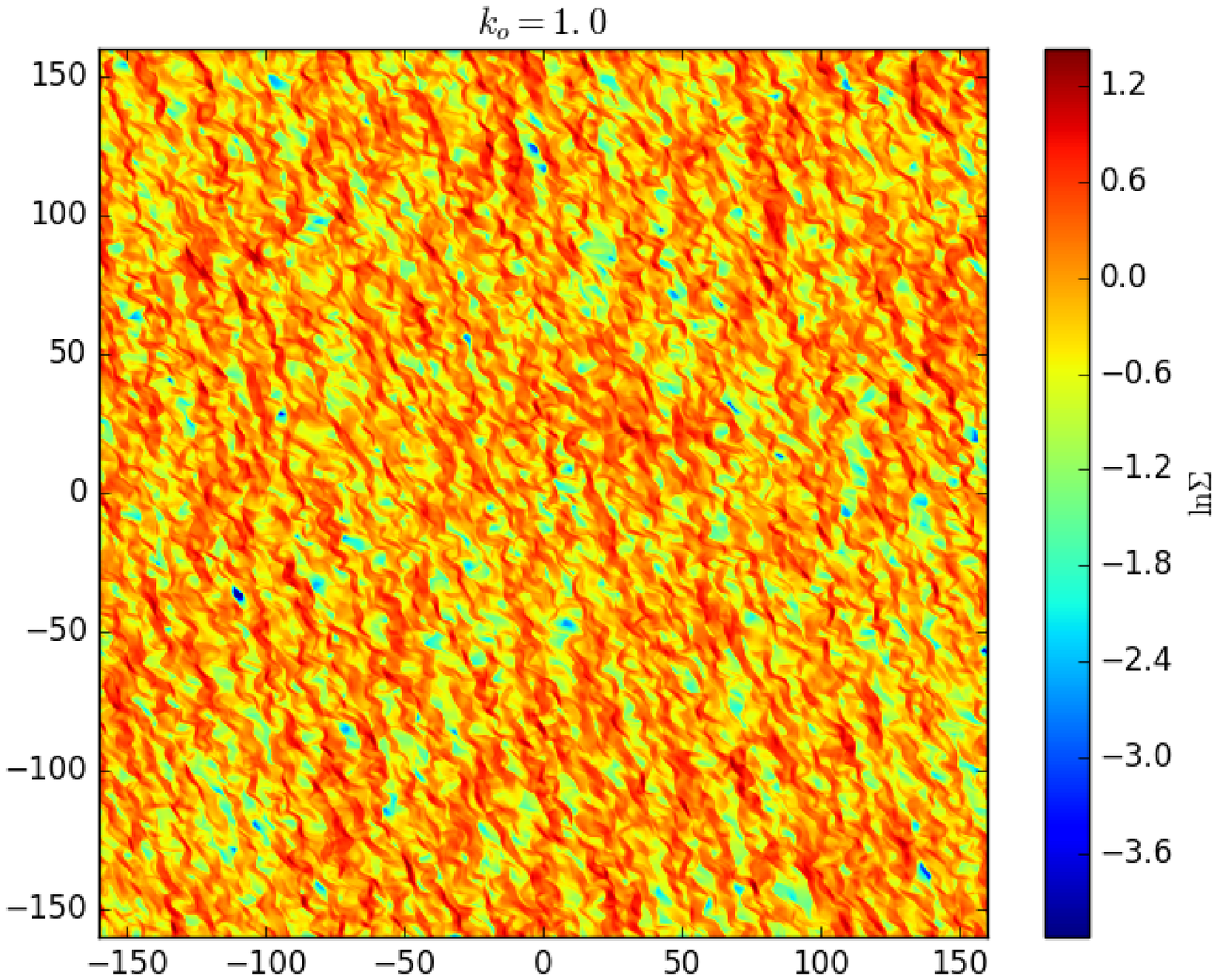}
\caption{Density structure for two simulations, one forced with a forcing wavenumber of $k = 0.1$ (top panel) and the other with a forcing wavenumber 
of $k = 1.0$ (bottom panel). Both simulations are isothermal and do not include
the self-gravity of the disc gas.  For $k = 0.1$, the forcing occurs at scales about one-fifth the size of the box and so the turbulence develops spiral
wave like features.  When forced on smaller scales ($k = 1.0$) obvious spiral like features do not develop.}
\label{fig:turbsheet}
\end{center}
\end{figure}

Figure \ref{fig:turbsheet} shows the density structure for two simulations, one forced with a forcing wavenumber of $k = 0.1$ (top panel) and the other 
with a forcing wavenumber of $k = 1.0$ (bottom panel). Both simulations are isothermal and do not include the self-gravity of the disc gas. For a 
forcing wavenumber of $k = 0.1$, the forcing occurs at scales $\lambda = 2\pi/k = 20 \pi$, which is about one-fifth the size of the box and so spiral wave-like features develop. For $k=1.0$, 
the forcing occurs at much smaller scales and no obvious spiral-like features develop. 

\begin{figure}
\begin{center}
\includegraphics[scale=0.45]{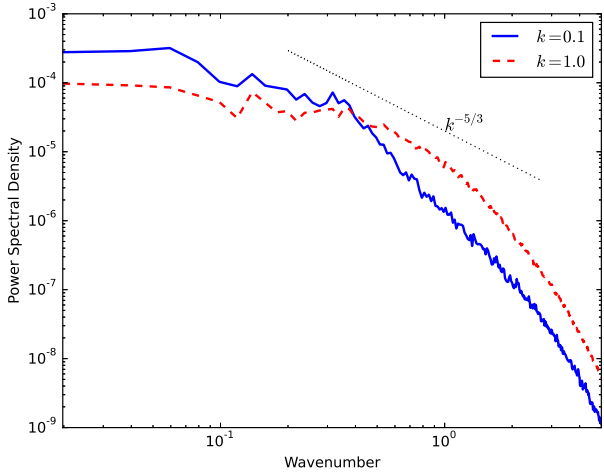}
\caption{Power spectra for the two simulations shown in Figure \ref{fig:turbsheet}. As expected, the simulation in which the turbulence is forced
at small scales ($k = 1.0$) has more power at these small scales than the simulation in which the turbulence is forced at larger scales ($k = 0.1$). A $k^{-5/3}$ power law (dotted line) is shown for reference.}
\label{fig:powerspectrum}
\end{center}
\end{figure}

Figure \ref{fig:powerspectrum} shows the power spectra for the two simulations shown in Figure \ref{fig:turbsheet}. As expected, the simulation
in which the turbulence is forced at small scales ($k = 1.0$) has more power at these small scales, than the simulation where the forcing occurs
at larger scales ($k = 0.1$). In both of these simulations, the turbulence reaches a quasi-steady state, in which the power spectrum remains approximately constant 
in time, about 10 timesteps after being turned on. A $k^{-5/3}$ power law (dotted line) is shown for
reference.

Table \ref{tab:turbsims} shows the properties of the various turbulence-only simulations (i.e., isothermal and with no self-gravity). The columns show
the forcing amplitude, $f_o$, the forcing wavenumber, $k$, and the resulting Mach number, $\mathcal{M}$. We estimate the Mach number
by determining the mean rms velocity once the simulation has settled into a quasi-steady state and then dividing this by the isothermal sound speed.  
What this indicates is that the turbulence can become supersonic if the forcing amplitude is sufficiently large ($f_o \sim 1$).

\begin{table}
\caption{Turbulent simulation parameters.}
\centering
\begin{tabular}{c c c}
\hline\hline
$f_o$ & $k$ & $\mathcal{M}$ \\
\hline
0.1   &  0.1   &  0.14 \\
0.1   &  1.0   &  0.27 \\
0.25  &  0.1   &  0.30 \\
0.25  &  1.0   &  0.57 \\
0.5   &  0.1   &  0.57 \\
0.5   &  1.0   &  0.95 \\
1.0   &  0.1   &  1.09 \\
1.0   &  1.0   &  1.5  \\
2.0   &  0.1   &  2.0  \\
2.0   &  1.0   &  2.32 \\
5.0   &  0.1   &  3.90 \\
\hline
\end{tabular}
\label{tab:turbsims}
\end{table}

\section{Results} \label{results}

\subsection{Baseline simulations} \label{base_sims}
\begin{figure}
\begin{center}
\includegraphics[scale=0.45]{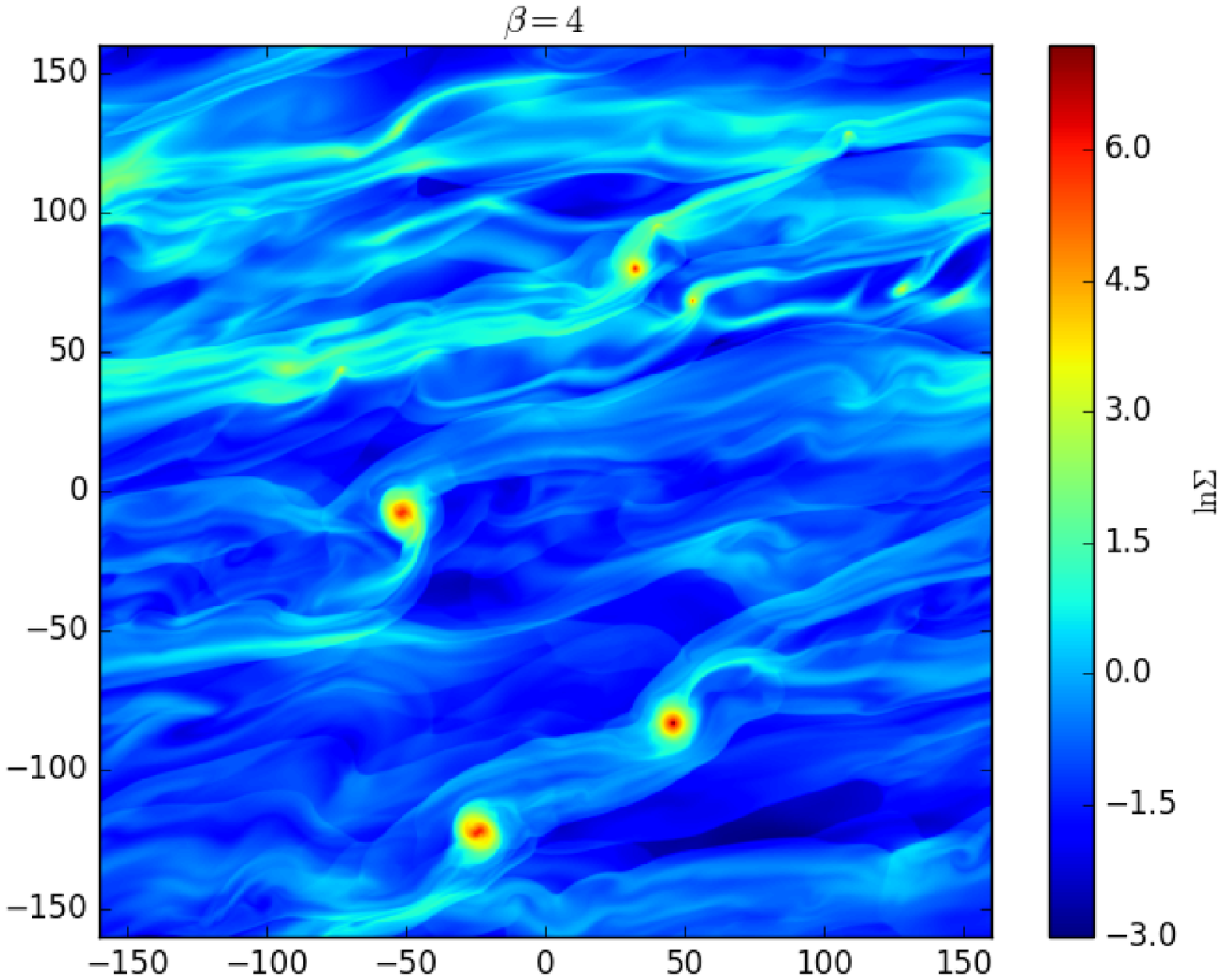}
\includegraphics[scale=0.45]{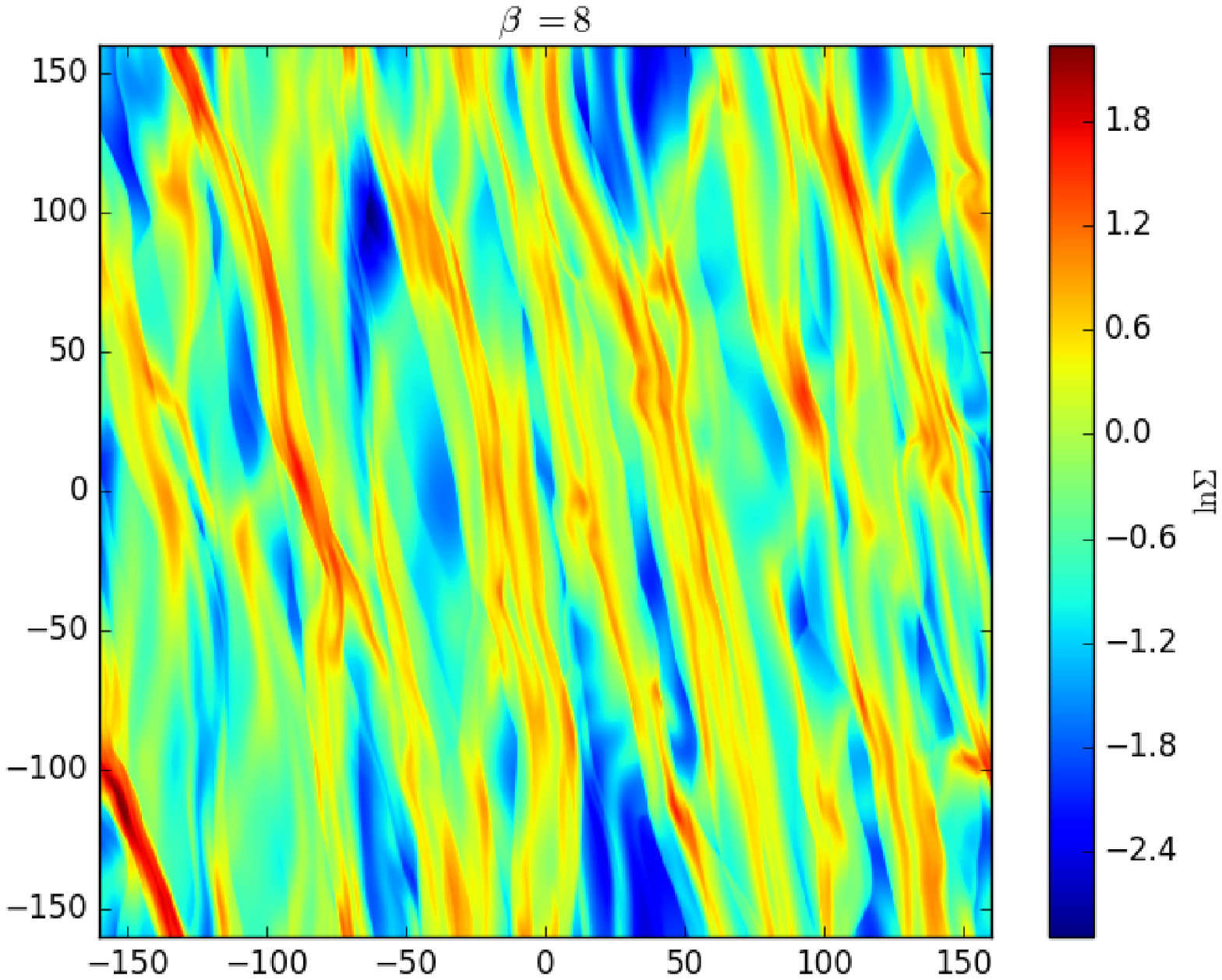}
\caption{Density structure for two simulations, one with a cooling time of $\beta = 4$ (top panel) and the other with a cooling time of $\beta = 8$ (bottom panel), but that do
not include additional turbulent forcing. The top panel shows a number of high-density clumps, indicating that this simulation is undergoing fragmentation.  The bottom shows spiral density waves, but no high density clumps, indicating that this simulation has settled into a quasi-steady state in which the instability is primarily acting to transport angular 
momentum.}
\label{fig:noturbsheet}
\end{center}
\end{figure}

We take our baseline simulations to be those initialised as described in Section \ref{init_conds} with an energy equation and cooling, but without any additional turbulent forcing.  In other words, the only turbulence is the gravito-turbulence driven by the gravitational instability itself.
Figure \ref{fig:noturbsheet} shows a baseline simulation with $\beta = 4$ (top panel) and one with $\beta = 8$ (bottom panel).  The $\beta = 4$ simulation has a number of dense 
clumps, indicating that it is undergoing fragmentation.  The $\beta = 8$ simulation does not, and instead shows the presence of spiral density waves, indicating that it is
in a quasi-steady state in which the instability is acting to steadily transport anglar momentum \citep{lodato04}. The fragmentation boundary in these simulations ($\beta \sim 4$)
is similar to that found by \citet{gammie01} ($\beta \le 3$), but we don't claim that ours is a precise representation of the fragmentation boundary. Table \ref{tab:basesims} shows the results
of our baseline simulations.  The $\beta = 4$ simulation shows clear signs of fragmentation, while the $\beta = 7, 8$ and $10$ simulations all settle into a quasi-steady state. 
The $\beta = 5$ simulation does have some regions of high density, but doesn't unequivocally fragment.  This could be an indication of stochastic fragmentation \citep{paardekooper12, young16}, or these
high density regions may simply shear out and the system maintain an approximately quasi-steady state.

\begin{table}
\caption{Baseline simulation results.}
\centering
\begin{tabular}{c c}
\hline\hline
$\beta$ & Disc fragments? \\
\hline
4   &  Yes   \\
5   &  ?   \\
7  & No      \\
8  &  No     \\
10   &  No    \\
\hline
\end{tabular}
\label{tab:basesims}
\end{table}

\begin{figure}
\begin{center}
\includegraphics[scale=0.45]{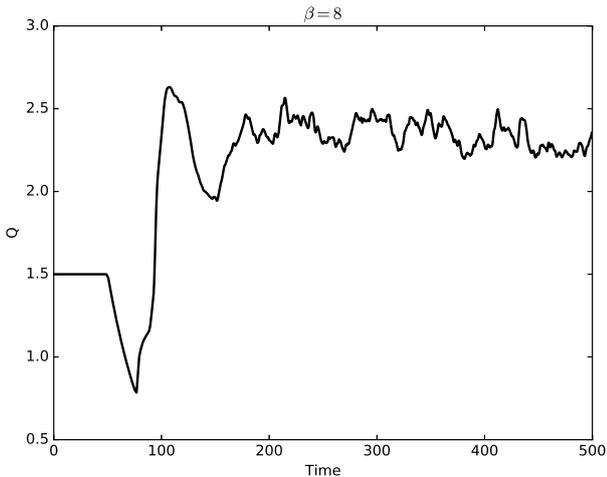}
\caption{Toomre $Q$ profile for the simulation, without additional turbulence, in which the cooling time is $\beta = 8$. The cooling is turned on at $t = 50$ and allowed
to decay to the prescribed rate over the next 50 timesteps, after which it is kept constant.  By $t \sim 200$ the system has settled into a quasi-steady state
in which $Q$ is roughly constant.} 
\label{fig:Q_beta8_F0}
\end{center}
\end{figure}

\begin{figure}
\begin{center}
\includegraphics[scale=0.45]{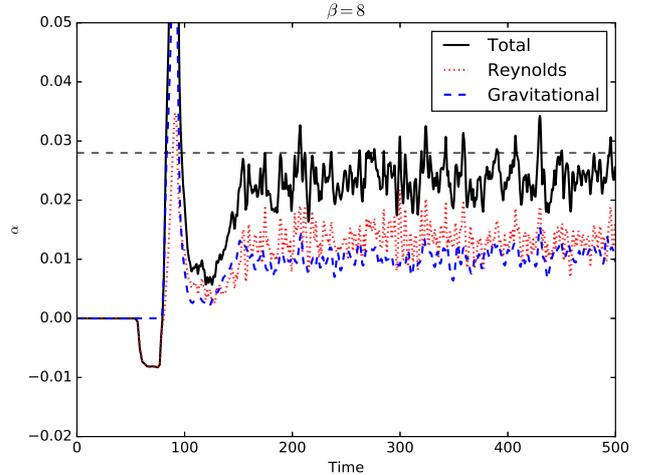}
\caption{Reynolds, gravitational and total stresses, expressed in terms of a viscous $\alpha$, in the simulation without additional turbulence and in which $\beta = 8$.  As with Figure \ref{fig:Q_beta8_F0},
the system settles into a quasi-steady state by $t \sim 200$. The dashed horizontal line
at $\alpha = 0.028$ shows the expected value of $\alpha$ based on the rate at which the effective viscosity
will dissipate energy matching the rate at which the system is cooling. What this figure also shows is that if the system does settle into a quasi-steady state, the instability
will act to transport angular momentum.}
\label{fig:alpha_beta8_F0}
\end{center}
\end{figure}

For the simulations that do not fragment, we can also consider other properties. Figure \ref{fig:Q_beta8_F0} shows the $Q$-profile for the $\beta = 8$ simulation.
This shows that we initially run the simulation, until $t = 50$, without any cooling. 
We then turn on the cooling and allow it to decay to the prescribed rate ($\beta = 8$) 
over the next 50 timesteps. By $t \sim 200$ the system settles to a state in which $Q$ is approximately constant.  This illustrates why, in simulations with additional turbulence,
we introduce the turbulence at $t = 200$; it is when - in the absence of additional turbulence - the system will typically have settled into its final state. Figure \ref{fig:alpha_beta8_F0} shows the stresses
in the $\beta = 8$ simulation. We express this in terms of the effective viscous $\alpha$ \citep{shakura73} and show the Reynolds, gravitational and total (Reynolds plus gravitational) stresses. 
Similarly to Figure \ref{fig:Q_beta8_F0} this shows how the system settles into a quasi-steady state by $t \sim 200$ and also illustrates how the instability is acting
to transport angular momentum. Figure \ref{fig:alpha_beta8_F0} also shows the $\alpha$ we would expect, given a cooling time of $\beta = 8$ (dashed line). This can be determined
by assuming that the rate at which an effective viscosity dissipates energy matches the rate at which the system is cooling \citep{gammie01,lodato04,rice05}, which gives
\begin{equation}
\alpha = \frac{4}{9 \gamma \left(\gamma - 1 \right) \beta}.
\label{eq:alpha}
\end{equation}
For the simulations that settle into a quasi-steady state ($\beta = 7, 8$ and $10$) the resulting quasi-steady total $\alpha$ is close to what would be expected based 
on Equation (\ref{eq:alpha}).

\subsection{Turbulence simulations}
We carried out a series of simulations using the same cooling times as in the baseline simulations (discussed in Section \ref{base_sims}) but in which we include
a turbulent forcing, as described in \S \ref{turb_forcing}, with a range of amplitudes (see Table \ref{tab:turbsims}) and with forcing wavenumbers of $k = 0.1$, and $k = 1.0$. As illustrated in
Table \ref{tab:turbsims} this means that we will be introducing turbulent forcings that vary from subsonic (for small forcing amplitudes, $f_o$) to supersonic (for
forcing amplitudes, $f_o$, that exceed about unity). We also consider three different regimes. A regime in which, according to the baseline simulations, fragmentation
will occur in the absence of an additional turbulent forcing, a regime close to the fragmentation boundary and in which high density regions may form but do not necessarily produce fragments, and a regime in which - in the absence 
of a turbulent forcing - the system settles into a quasi-steady, non-fragmenting state.

\subsubsection{Baseline fragmenting case} \label{basefrag}

In these simulations (with $\beta = 4$) the system undergoes fragmentation in the absence of a turbulent forcing. Here we consider what happens if we then add
a turbulent forcing with wavenumbers of $k = 0.1$ and $k = 1.0$ and with forcing amplitudes, $f_o$, that vary from $0.1$ to $1.0$. For
small forcing amplitudes ($f_o = 0.1$) there is no difference between the baseline simulation and the turbulently forced simulation. For turbulent forcings 
with $k = 0.1$ there is also little difference between the baseline simulations and the turbulently forced simulations, for all forcing amplitudes; they all fragment. 

\begin{figure}
\begin{center}
\includegraphics[scale=0.45]{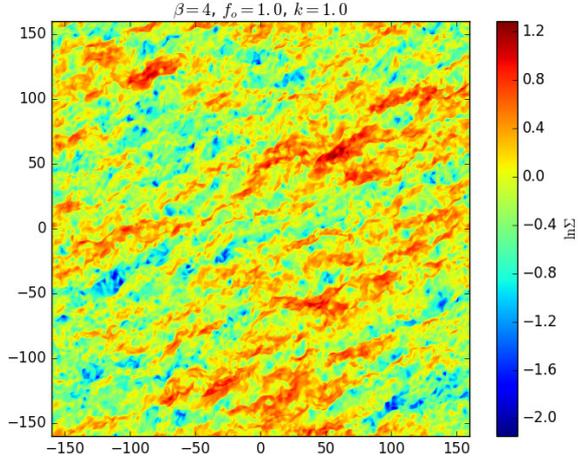}
\caption{Figure showing the surface density structure in a simulation with $\beta = 4$ and with an imposed turbulence (initiated at $t = 200$) with $k = 1.0$,
and $f_o = 1.0$. In the absence of this imposed turbulence, this simulation fragments (top panel of Figure \ref{fig:noturbsheet}).  Turbulence forced at $k = 1.0$ appears, however, to inhibit fragmentation.}
\label{fig:beta4_F1_k1}
\end{center}
\end{figure}

\begin{figure}
\begin{center}
\includegraphics[scale=0.45]{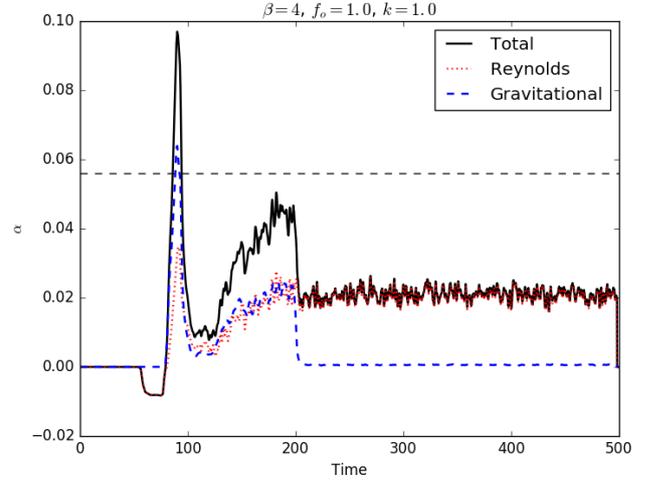}
\caption{The Reynolds, gravitational and total (Reynolds plus gravitational) stresses for the simulation shown in Figure \ref{fig:beta4_F1_k1}. The horizontal dashed line at 
$\alpha = 0.056$ shows the value of
$\alpha$ to which we'd expect the system to tend, given that the cooling time is $\beta = 4$. The total $\alpha$ (black line) appears to be tending towards the expected value 
until $t = 200$, at which point the imposed turbulence is initiated. At this point, the total $\alpha$ drops and becomes dominated by the Reynolds stress.  The gravitational
$\alpha$ virtually disappears. The system has settled into a state of thermal equilibrium in which the imposed cooling is balanced by the dissipation of the Reynolds stress and
dissipation of the imposed turbulence, and with a very weak gravitational instability.}
\label{fig:alpha_beta4_F1_k1}
\end{center}
\end{figure}

However, for $k = 1.0$ there appears to be a forcing amplitude above which the turbulent forcing inhibits fragmentation.  Figure \ref{fig:beta4_F1_k1} shows
a simulation with $\beta = 4$, $f_o = 1.0$, and $k = 1.0$. There are no indications of any high density fragments. Figure \ref{fig:alpha_beta4_F1_k1} shows
the effective viscous $\alpha$ values for this simulation. For $\beta = 4$, we would expect the total $\alpha$ to settle somewhere close to $\alpha = 0.056$. Prior to turning
on the turbulent forcing at $t = 200$, the total $\alpha$ is heading towards the expected value, but drops once the turbulent forcing is initiated. The Reynolds contribution continues
to provide about half of the expected stress, but the gravitational contribution virtually disappears. 

What appears to be happening is that the imposed turbulence
is non-helical and, hence, does not necessarily manifest itself as a stress and does not necessarily transport angular momentum.  It does, however, heat the disc, which can be seen
by considering $Q$, which increases from just over $2$ to about $5$. The system therefore settles into a state of thermal equilibrium, with the imposed cooling
being balanced by dissipation via the Reynolds stresses and via dissipation of the imposed turbulence. The gravitational instability becomes very weak,
and fragmentation is suppressed. The weakness of the gravitational stresses suggests that the Reynolds stresses are being driven by the imposed turbulence, 
possibly through coupling into the spiral modes  (see, for example, \citealt{heinemann09} and \citealt{mamatsashvili11}).

\subsubsection{Baseline boundary case}\label{basenear}

We also consider how an additional turbulence influences simulations near the fragmentation boundary, in this case those with cooling times of $\beta = 5$ and $\beta = 7$.  In the baseline simulation,
some high density regions did form in the $\beta = 5$ simulation, but by the end of the simulation ($t = 500$) there was not unequivocal indications of fragmentation. In the $\beta = 7$ simulation there
were no high-density regions by the end of the simulation, and the system appears to have settled in a quasi-steady, non-fragmenting state.  We again impose turbulent forcings 
with forcing wavenumbers of $k = 0.1$ and $k = 1.0$ and forcing amplitudes, $f_o$, between $0.1$ and $1.0$. Again, for low forcing amplitudes 
($f_o = 0.1$) there is little difference between the baseline simulations and the turbulently forced simulations.

\begin{figure}
\begin{center}
\includegraphics[scale=0.45]{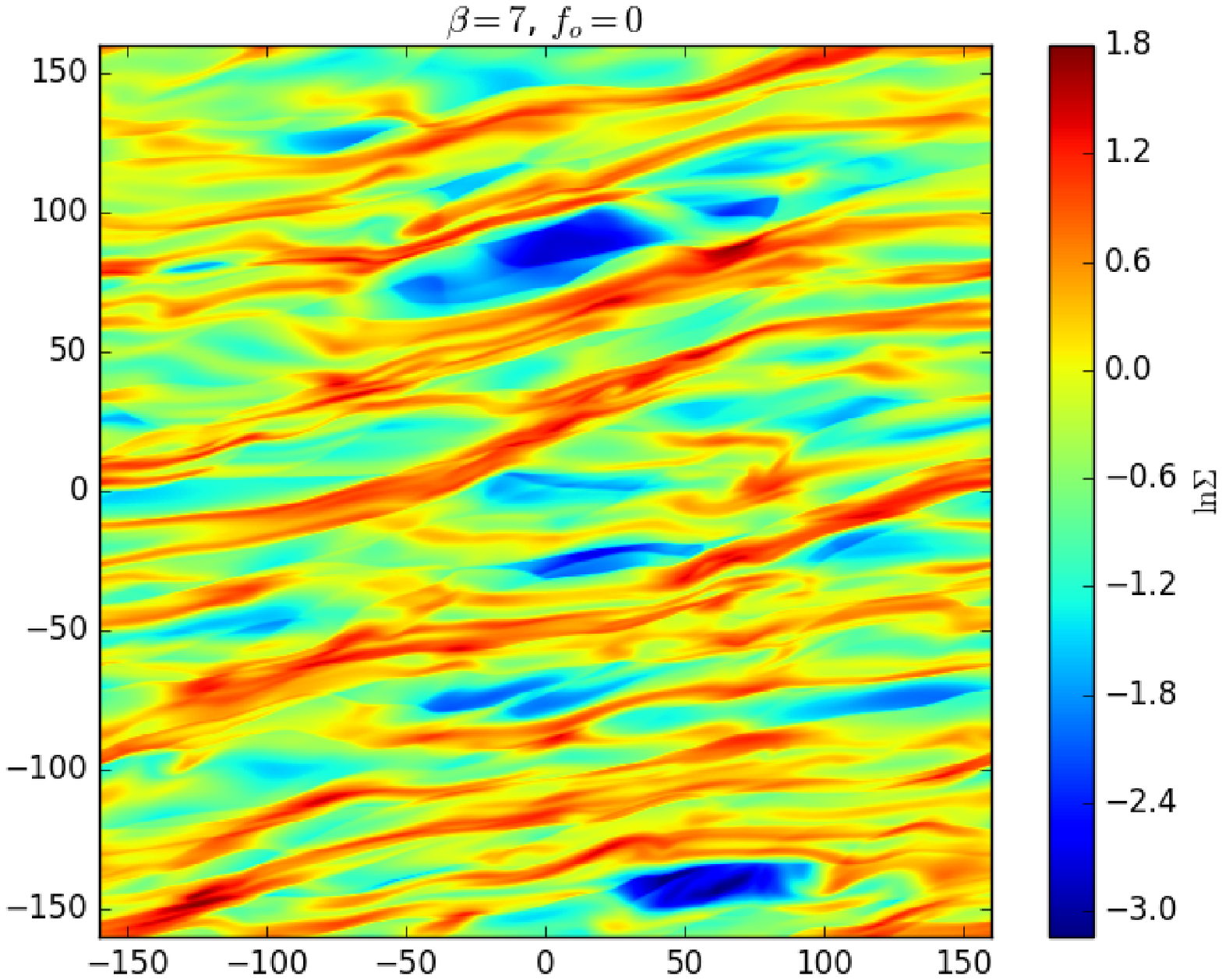}
\includegraphics[scale=0.45]{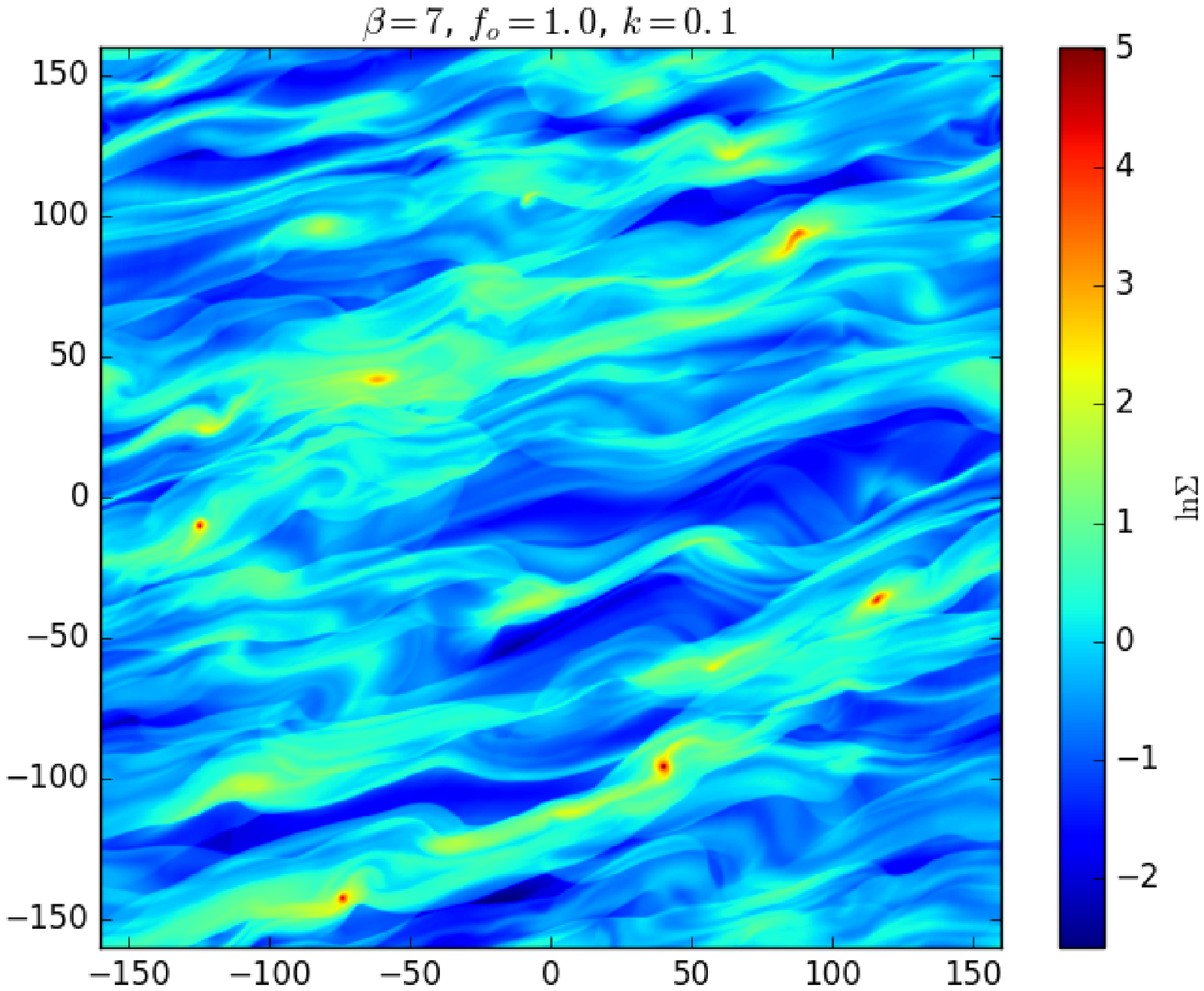}
\caption{Figure showing the baseline simulation for a cooling time of $\beta = 7$ (top panel) and the simulation with the same cooling time but that also includes turbulent forcing,
forced at a wavenumber of $k = 0.1$ and with a forcing amplitude of $f_o = 1.0$ (bottom panel). The baseline simulation shows no signs of fragmentation, while the simulation
with turbulent forcing does.  This suggests that it might be possible for imposed turbulence to trigger fragmentation if the system is already quite close to the fragmentation boundary
and if the turbulence is forced at small enough wavenumbers.}
\label{fig:beta7_comp}
\end{center}
\end{figure}

However, in this case as we increase the forcing amplitude, those forced at wavenumbers of $k = 0.1$ start to undergo fragmentation. 
This is illustrated in Figure \ref{fig:beta7_comp} which shows
the baseline simulation with $\beta = 7$ (top panel) and the simulation with the same cooling time, and at the same simulation time, 
but forced with turbulence with wavenumber $k = 0.1$ and amplitude $f_o$ = 1.0. The top panel shows that the baseline simulation settled into a quasi-steady,
non-fragmenting state, while the bottom panel indicates that the simulation with turbulent forcing ($k = 0.1$, $f_o = 1.0$) has fragmented. We found a similar result with the $\beta = 5$ simulation and that this appeared to initially fragment at slightly lower forcing amplitudes ($f_o = 0.5$) than the $\beta = 7$ simulation. 

However, as a final test, we increased the turbulent forcing amplitude to $f_o = 2$, while keeping $k=0.1$.  This also led to fragmentation in the $\beta = 5$ case, but not in the $\beta = 7$ case (which did fragment for $f_o = 1$).   In the latter case, the additional turbulence acted to heat the disc, increasing $Q$, and - ultimately - making it more stable. 

In contrast to the simulations forced at a wavenumber of $k = 0.1$, those forced with larger wavenumbers ($k = 1.0$) behave as described in Section \ref{basefrag}; they settle into 
quasi-steady states in which fragmentation is inhibited and in which the instability is very weak.  

This suggests that turbulent forcing might promote fragmentation if the system is already near the fragmentation
boundary and if the forcing is not at scales that ultimately disrupt the fragments and predominantly heats the disc, driving it towards becoming more stable.  Fragmentation therefore appears to occur if the forcing is at scales that allow the turbulence to couple into the spiral modes and, hence, does not necessarily disrupt the gravitational instability, while also not being of such a high amplitude that heating of the disc dominates to such an extent that the instability is largely suppressed.

\subsubsection{Baseline quasi-steady case} \label{basesteady} 

\begin{figure}
\begin{center}
\includegraphics[scale=0.45]{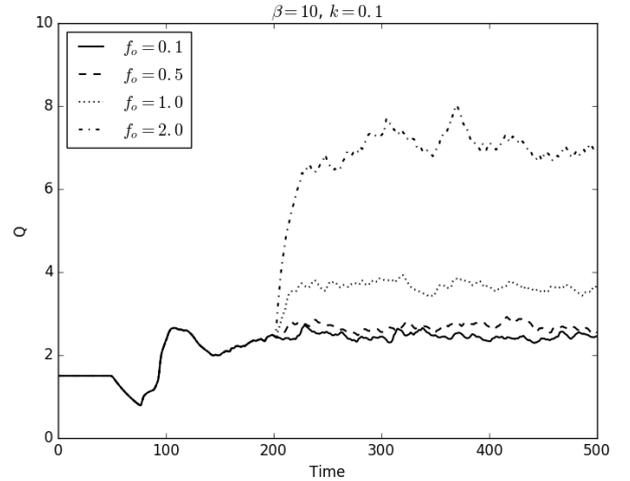}
\caption{$Q$ values from a set of simulations with cooling times of $\beta = 10$ and with various turbulent forcing amplitudes, but all forced at a wavenumber of $k = 0.1$. They all settle to
quasi-steady states with constant $Q$, but those with larger forcing amplitudes settle to states with higher $Q$ values and are, therefore, more gravitationally stable.}
\label{fig:Q_beta10_comp_k01}
\end{center}
\end{figure}

In this case we introduce an additional turbulent forcing into baseline simulations that show no signs of fragmentation and that settle into a quasi-steady state, i.e., those with $\beta = 8$ and $\beta = 10$. In these simulations,
the system remains quasi-steady and the turbulence acts to heat the disc and make it more stable, rather than promoting fragmentation. For example, Figure \ref{fig:Q_beta10_comp_k01} shows the $Q$ values
for simulations with the dimensionless cooling times of $\beta = 10$ and with various turbulent forcing amplitudes, all forced at a wavenumber of $k = 0.1$. They all settle to quasi-steady states with constant $Q$,
but those with larger forcing amplitudes settle to states with larger $Q$ values and are, therefore, more gravitationally stable.  We see the same effect if the turbulence is forced at wavenumbers
of $k = 1.0$.

\begin{figure}
\begin{center}
\includegraphics[scale=0.45]{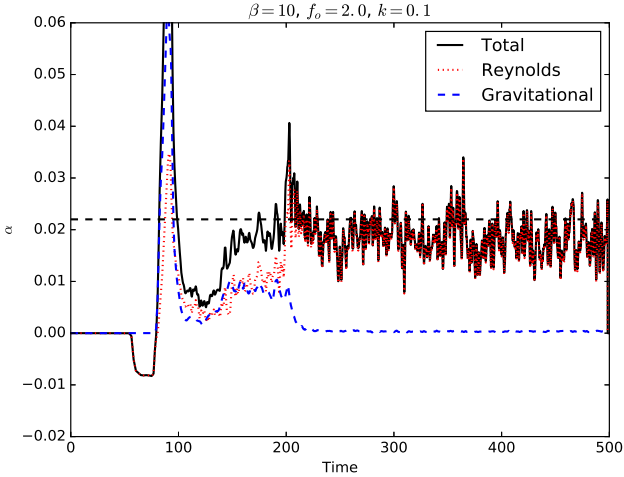}
\caption{The Reynolds, gravitational and total (Reynolds plus gravitational) stresses for the simulation with $\beta = 10$ and with turbulence forced at a wavenumber $k = 0.1$ and with
an amplitude of $f_o = 2.0$. The horizontal dashed line at $\alpha = 0.022$ shows the expected value of $\alpha$ given that $\beta = 10$. When the turbulence is initiated ($t = 200$) the gravitational stress virtually disappears, and the system essentially becomes gravitationally stable, with
the imposed cooling balanced by dissipation of the Reynolds stresses and the dissipation of the imposed turbulence.  In this case, unlike the case shown in Figure \ref{fig:alpha_beta8_F0},
the Reynolds stress is quite close to what would be expected from energy balance estimates (dashed line), suggesting that the imposed turbulence has strongly coupled into the spiral modes, which then produces
Reynolds stresses that would also act to transport angular momentum.}
\label{fig:alpha_beta10_F2_k01}
\end{center}
\end{figure}

Firgure \ref{fig:alpha_beta10_F2_k01} shows the stresses, as represented by the effective viscous $\alpha$ for simulation with $\beta = 10$ and with turbulence forced at a wavenumber of $k = 0.1$
and with a forcing amplitude of $f_o = 2.0$. As shown in Figure \ref{fig:Q_beta10_comp_k01} the turbulence heats the disc, making it more gravitationally stable. What Figure \ref{fig:alpha_beta10_F2_k01} shows is that this
also depresses the gravitational stress, so that - in a quasi-steady state - the cooling is balanced by dissipation due to the Reynolds stresses and dissipation of the imposed turbulence. Unlike 
Figure \ref{fig:alpha_beta8_F0}, Figure \ref{fig:alpha_beta10_F2_k01} shows that the Reynolds stresses end up quite close to what would be expected based on the imposed cooling (dashed line).  This suggests that 
there are cases in which the imposed turbulence can strongly couple into the spiral modes, which then produces Reynolds stresses that also act to transport angular momentum \citep{heinemann09,mamatsashvili11}. 

\section{Instantaneous velocity kick experiments} \label{sec:velkick}

We now describe experiments in which a marginally stable self-gravitating disc is affected by an instantaneously applied turbulent field velocity kick. These experiments are complimentary to those presented earlier where turbulence was continuously driven by a force field. In the case of instantaneous velocity kicks we gain a direct insight into тче temporal evolution of the dense regions formed by convergent turbulent velocity field.

\subsection{Numerical setup for SPH experiments}

\subsubsection{The code and initial conditions}

We again utilize the popular ``$\beta$-cooling" model where the local cooling time of the gas is given by $\tau_{\rm c} = \beta \Omega^{-1}$ (see eq. \ref{eq:energeq}). We use a Smoothed Particle Hydrodynamics (SPH) code Gadget 3 \citep{Springel05} and set up a gas disc of initial mass $M_{\rm d} = 0.3 \msun$ in orbit around 1 solar mass star. The disc inner and outer radii are 20 AU and 200 AU, respectively. The initial surface density profile is given by $\Sigma(R) \propto 1/R$ with the normalisation set by the total disc mass $M_{\rm d}$. The disc is initially hot and vertically extended so that the Toomre parameter is well above unity everywhere. The disc is then relaxed (e.g., simulated with no velocity kicks applied) for $\sim 10$ orbits on the outer edge. 

Depending on the value of $\beta$ cooling parameter, the disc then either fragments or settles into a self-regulated gravito-turbulent state. In these 3D experiments, we use adiabatic index $\gamma=5/3$  which is more appropriate for us here since we work in 3 dimensions here rather than 2 as in Sections \ref{Method}-\ref{results} \citep[see also section 2 in][]{gammie01}. For the total initial number of SPH particles $N = 0.8$ Million, we find that our disc fragments for $\beta \le 8$ and does not fragment if $\beta > 9$. Note that the fragmentation boundary is resolution  and artificial viscosity prescription dependent, as discussed in \S \ref{Method}. However, analytical arguments presented in the Discussion section show that the main results should not depend significantly on the exact value of the critical value of $\beta$ as long as it is well above unity. 

To test whether stochastic compression caused by convergent turbulent velocity flows helps to promote disc fragmentation, we chose the disc with $\beta =10$ as our initial condition. This value of $\beta$ is just above the fragmentation boundary for our setup and hence even a mild amount of turbulent compression may be expected to affect the results strongly if the model of \cite{hopkins13} is correct.

\subsubsection{Turbulent velocity kick field}

We apply velocity kicks only in the directions along the disc midplane, neglecting the velocity kicks in the vertical ($z$) direction. This is done to avoid modelling ambiguities in cases when the largest turbulent scales exceed the disc scale height, $H$. This choice should not affect the overall outcome of our experiments. We shall later find that turbulent velocity kicks applied just along the disc midplane do create high density regions, and so may be expected to aid disc collapse.

We first setup a turbulent three-dimensional velocity field for isotropic turbulence with the Kolmogorov power spectrum in a cube with size equal to $2L_b = 1$. The calculation is performed in a uniform 3D grid that fills the cube. We then pick the turbulent velocity field $(v^t_x(x,y), v^t_y(x,y))$ along the $z=0$ plane of the cube and map it to velocity kicks $\mathbf{\Delta v}$ in the disc. To enable this mapping, we relate the disc coordinates along the midplane to the $(x,y)$ coordinates in the cube by a linear transformation:

\be
x = x' \frac{L_b}{R_0}\qquad {\rm and}\qquad y = y' \frac{L_b}{R_0}\;,
\label{eq:transf}
\ee
where $x'$ and $y'$ are the coordinates in the disc and $R_0 = 150$ AU. The turbulent velocity kicks passed to the SPH particles are then scaled to the local Keplerian velocity, $v_{\rm K}$,
\be
\mathbf{\Delta v} = K \mathbf{v^t} v_{\rm K}\;
\label{eq:dv}
\ee
where $K$ is a dimensionless parameter. This scaling makes physical sense as the turbulent velocities are expected to scale with the local sound speed, which in turn usually scales approximately linearly with $v_k$ for discs beyond tens of AU. We select the value of $K$ such that it would result in a desired value for the Mach number for turbulent kicks, defined as
\be
M_{\rm t} \equiv \overline{\left(\frac{\Delta v^2} {c_s^2}\right)^{1/2}}\;,
\ee
where the line over the right hand side means averaging over all SPH particles in the disc region where the kicks are applied (disc radius less than $R_0$).

\subsection{Results}\label{sec:res}

\begin{figure*}
\includegraphics[width=0.45\textwidth]{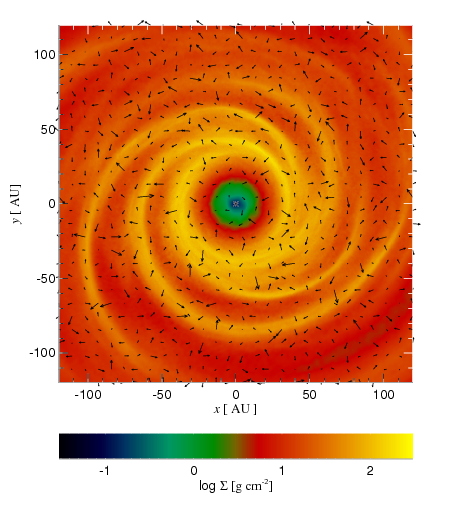}
\includegraphics[width=0.45\textwidth]{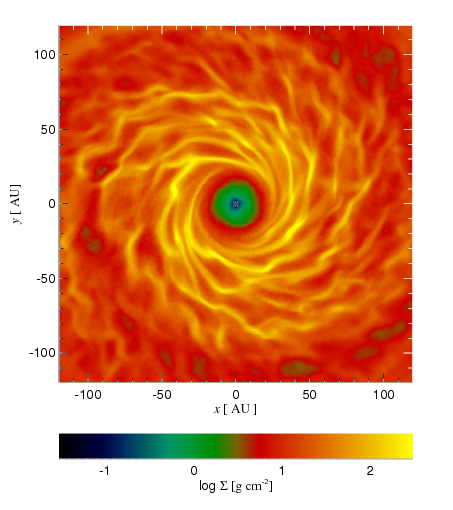}
\includegraphics[width=0.45\textwidth]{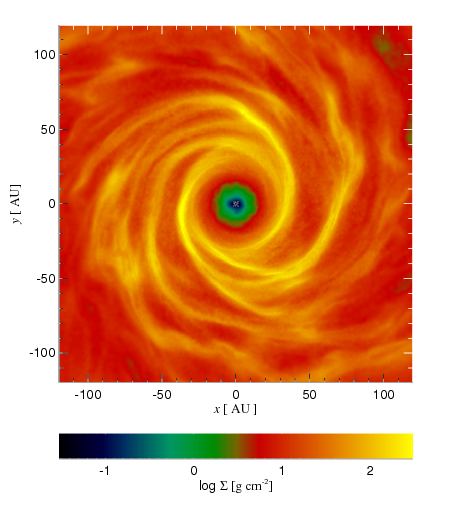}
\includegraphics[width=0.45\textwidth]{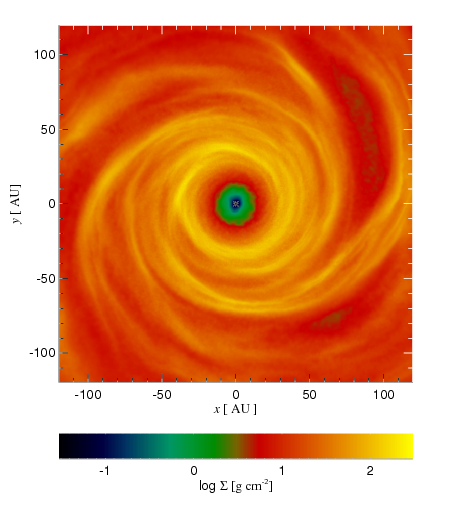}
\caption{Disc surface density maps for turbulent Mach number $M_{\rm t} =1$ instantaneous velocity kick
experiment. The top left panel, corresponding to $t=0$, also shows the map of the turbulent velocity kick, $\mathbf{v^t}$. The corresponding times for the panels are $t=0.5$, 2 and 5 in code units. }
 \label{fig:SPH_M1}
 \end{figure*}

Figure \ref{fig:SPH_M1} shows the disc surface density map at four different times for the simulation in which the turbulent Mach number $M_{\rm t}$ is set to 1. The top left panel shows the initial condition, with the map of the velocity kick $\mathbf{v_t}$ super-imposed. Although the latter quantity is not exactly the velocity kick $\mathbf{\Delta v}$, which is further scaled by the local velocity field (cf. eq. \ref{eq:dv}), we show it for clarity purposes as it is uniform across the figure whereas the velocity kicks decrease with $R$ and are hence less discernible at large radii.

Note that the velocity kick field shows structure on scales $\le H \approx 0.15 R$, where $R$ is the local disc radius, as expected from locally driven turbulence \citep{hopkins13}. There are many regions of convergent velocity flows in the panel.

The next panels show times $t=0.5$, 2 and 5 in code units, where the time code unit is $1/\Omega(R=100 {\rm AU}) \approx 160$ years in physical units. We observe that the dispersion in density across the disc increases significantly at $t=0.5$, and that the highest projected column densities are now much greater than they were at $t=0$. However, disc column density plots at later times show that these local density increases are transient. The panel with $t=5$ in particular shows that none of the high density regions survive after about one revolution (which takes in code time units $2 \pi$ at radius $R=100$ AU).

There is another useful way of analysing the gas density evolution. Define the normalised gas density, 
\be
\rho_{\rm norm} = \frac{\rho}{\rho_{\rm tid}}\;,
\label{rho_norm}
\ee
where $\rho_{\rm tid}$ is the tidal density,
\be
\rho_{\rm tid} = \frac{M_*}{2\pi R^3}\;.
\ee
For gravitational collapse to occur, we need $\rho_{\rm norm} \gg 1$ or else the region is sheared away by tidal forces from the central star. Additionally, the region needs to be large/massive enough to be self-gravitating.

\begin{figure}
\includegraphics[width=0.5\textwidth]{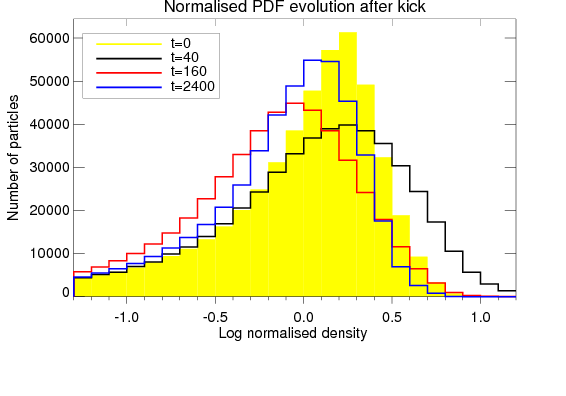}
\caption{The histogram of $\log \rho_{\rm norm}$ as defined by equation \ref{rho_norm} for SPH particles at several different times for the simulation with Mach number $M_{\rm t} =2$, as described in \S \ref{sec:res}. The relevant times are shown in units of years in the legend. Note that the high density tail of the histogram is amplified quickly after the turbulent kicks are applied but then disappears after about one dynamical time scale ($\sim 160$~years). This shows that the high density structures are not self-bound; they are transient regions that decompress quickly when turbulence decays away.}
 \label{fig:PDF}
 \end{figure}

In Figure \ref{fig:PDF} we plot the Probability Density Function (PDF) for the normalised density defined on SPH particles before the kicks are applied (the histogram shaded with the yellow colour), and after the kicks at several different times given in the legend in years. For this simulation the turbulent kicks are stronger, $M_{\rm t} = 2$. At $t=0$, there is a high density tail of particles with $\rho_{\rm norm} \gg 1$ but we note that the disc is stable on arbitrarily long time scales, implying that these high density regions are not massive enough to be self-gravitating.

At time $t = 40$ years ($t=0.25 $ in code units, the black histogram), the high density tail of the particles is significantly more populated than at $t=0$, as one would expect, indicating that turbulent compression by converging flows does take place. However, after one dynamical time, $t=160$ years ($t=1$ in code units), these high density regions dissipate away (see the red curve), and it is the low density tail of the PDF that is now more populated. This again shows that convergent flows do not create long lasting self-bound density structures and that eventually the density of the disc in fact decreases due to the extra heat injected by the turbulent motions into the disc. The extra heat is dissipated on time scales of several cooling times, and the PDF eventually returns to the one before the kick (compare the blue and the yellow histograms).

To sample the parameter space, we varied the turbulent Mach number, sampling $M_{\rm t} = 0.2$, 0.5, 1, and 2, values. None of these tests yielded gravitational collapse of the $\beta = 10$ disc. 

We also varied the minimum wave number in the power spectrum of turbulence, investigating modes with wavelength comparable to $R$, and again ran same range of $M_{\rm t}$. The results of these experiments were very similar to those presented above -- none of these cases yielded collapse of the disc.

\section{Discussion}\label{discussion}

In this paper we tested the supersonic turbulence fragmentation theory for protoplanetary discs proposed by \cite{hopkins13} by means of numerical experiments. The theory is essentially an extension of the turbulence-regulated star formation theory worked out by \cite{krumholz05} for much larger scale discs -- galactic discs and discs in ultra-luminous infrared galaxies. The main premise of the theory is that turbulence creates a log-normal PDF distribution of gas densities within the disc, and that while most of the gas is of too low density, there are sufficiently high density regions in the high density tail of PDF that are gravitationally unstable and therefore can collapse.

However, there are significant differences in the physics of protoplanetary and galactic scale discs which render one-to-one transfer of knowledge from the latter to the former questionable. First of all, non-fragmenting protoplanetary discs cool slowly in terms of the local dynamical time, $\tau_c \gg 1$ \citep{gammie01}, whereas larger scale discs have very short cooling times, $\tau_c\ll 1$. Secondly, fragmentation of galactic scale discs is known to produce very energetic feedback via supernova explosions and winds from massive stars. These feedback processes are believed to be important in maintaining the turbulence and in keeping the discs from a very rapid -- dynamical -- fragmentation into stars \citep{thompson05}. In contrast, collapsing regions in protoplanetary discs are of planetary to brown dwarf mass range and are thus not able to produce explosive supernovae or radiation-pressure driven winds. Clumps in protoplanetary discs do produce radiative feedback on the surrounding gas, but this is only important for the Hill sphere region around the clumps \citep{NayakshinCha13,stamatellos15} and merely makes the gas somewhat hotter, rather than inducing supersonic turbulent motions.

To explore the effects of turbulence on protoplanetary discs, we performed numerical experiments of two kinds in this paper. In \S \ref{Method}  and \ref{results}, we used 2D grid based method with an imposed turbulent driving force to generate turbulent velocity fluctuations in a patch of the disc being simulated. It was found that turbulent driving does not generally increases the likelihood of disc fragmentation, and instead makes the discs more stable to fragmentation. In particular, \cite{hopkins13} suggested that discs far away from the fragmentation boundary (marked by the critical value of the dimensionless cooling time parameter $\beta_{\rm crit}\approx 5$, so discs with $\beta \gg \beta_{\rm crit}$) can nevertheless fragment due to stochastic turbulent fluctuations in density. We find that for $\beta$ well above $\beta_{\rm crit}$, the disc becomes {\em less} unstable in the presence of imposed turbulence (\S \ref{basesteady}), in contrast to the theory. Furthermore, even for $\beta < \beta_{\rm crit}$, when non-turbulent discs are expected to fragment vigorously, our simulated  discs become stable when the turbulent driving is large enough and is forced at a large enough wavenumber (\S \ref{basefrag}). Only near the fragmentation boundary, that is, at $\beta$ just above $\beta_{\rm crit}$, do we find some cases when the turbulent disc is more unstable than its counterpart without driven turbulence (\S \ref{basenear}). However, in the last case, there seems to be little practical importance to this result since the disc would become unstable if the value of $\beta$ is lowered slightly; just enough to fall below the fragmentation boundary. This may happen naturally in real discs if they were to gain some mass from the envelope, if their opacity were to fall due to grain growth, or if the disc radius increases. Certainly, the most intriguing result of \cite{hopkins13} -- disc fragmentation far away from what is usually consider the marginally stable state -- is not supported by our simulations.

In the second type of numerical experiments (\S \ref{sec:velkick}), we give instantaneous turbulent velocity kicks to a protoplanetary disc in a marginally stable state with $\beta$ slightly above the fragmentation boundary. The disc is simulated via a global SPH (particle based) method in this setting. Similarly to \S \ref{results}, it is found that the disc becomes more, rather than less, stable after the kicks. Although it is true that higher density regions do form in regions of convergent velocity kicks (cf. fig. \ref{fig:SPH_M1}), these regions are not self-bound and dissipate in several dynamical times. Analysis of the PDF distribution after the kick shows that after the dissipation of the high density regions, the high density tail of the PDF actually gets more depleted than before the kicks (fig. \ref{fig:PDF}), again showing that the disc becomes less, rather then more, unstable to fragmentation.

We think that the main reason why we get results different from the analytical theory of \cite{hopkins13} is that the said theory assumes isothermality, at least in the local sense, for the disc. This assumption may well be valid for rapidly cooling discs (i.e., when $\beta\ll 1$) which is indeed the case for large scale galactic discs and also for sufficiently hot discs in the ultra-luminous infrared galaxies \citep[see the Appendix in][]{thompson05}. Protoplanetary discs are however expected to be in the $\beta\gg 1$ regime near the fragmentation boundary \citep{gammie01}. In this situation, assuming that the disc Toomre parameter $Q$ and the turbulence strength, described for example by the mean Mach number of turbulent velocity structures ($\cal{M}$), are independent of one another, as the \cite{hopkins13} theory posits, is inconsistent. Indeed, $Q = c_s \Omega/\pi G\Sigma$, and we now show that both $Q$ and $c_s$ must increase if turbulence is local on scale $H$ and supersonic in a protoplanetary disc. 

Neglecting for a moment heating due to gravito-turbulence in the disc, the thermal balance in the disc implies that 
\be 
\frac{c_{\rm s}^2}{\beta}\; \Omega \gtrsim \frac{v_{\rm t}^2}{t_{\rm dec}}
\ee
where $t_{\rm dec}$ is the time scale for turbulence decay and $v_{\rm t} = {\cal M} c_{\rm s0}$ is the turbulent velocity expressed in relation to $c_{\rm s0}$, the sound speed in the disc before the turbulent driving is switched on. In this equation, the left hand side is the approximate cooling rate per unit mass of the gas, whereas the right hand side is the turbulent heating. Local turbulence with the largest eddy lengh scale $\lambda_{\rm t} \lesssim H$ will decay on time scale $t_{\rm dec} \sim \lambda_{\rm t}/v_{\rm t} \lesssim 1/(\Omega {\cal M})$. Thus, the sound speed in the disc heated by supersonic local turbulence needs to be at least
\be 
c_{\rm s} \gtrsim c_{\rm s0} \left( \beta {\cal M}\right)^{1/2}\;,
\label{cs_turb0}
\ee
to balance the imposed turbulent heating, which is significantly larger than the unperturbed sound speed for $\beta >1$ and ${\cal M} > 1$. Evidence for increase in the sound speed with the strength of the driven turbulence can be seen in fig. \ref{fig:Q_beta10_comp_k01} which shows the disc Toomre parameter versus time. 

This discussion shows that local supersonic turbulence produces more heat than a disc could radiate away if its cooling time parameter $\beta \gg 1$. Therefore increasing the turbulent ${\cal M}$ heats the disc and increases the $Q$ parameter, thus making the disc stable. In fact, for a fixed turbulent forcing amplitude, $f_o$, the increase in the disc sound speed may render the driven turbulence subsonic relative to the new sound speed, even if it is initially supersonic. For example, in driven turbulence simulation with $f_o = 2$, and $k_0 = 0.1$, the Mach number of turbulent motions is measured at ${\cal M} \approx 2$ if the disc is assumed to be isothermal. In a similar setup, but with the $\beta$-cooling in place (using $\beta = 10$) the disc heats up, and the turbulent motions in equilibrium result in a Mach number of only ${\cal M} = 0.55$. This is approximately consistent with equation \ref{cs_turb0}, showing that the sound speed of the disc increases by about the expected factor, $(\beta {\cal M})^{1/2} = \sqrt{20} = 4.5$.

On the other hand, in $\beta \ll 1$ discs, gas heating due to decaying turbulence can be significantly less than the gas radiative cooling rate, and this is why, in the fast-cooling regime, systems could be dominated by turbulent motions rather than thermal pressure support \citep{padoan02,krumholz05}. The argument made above, however, suggest that supersonic turbulence cannot even exist in slowly cooling protoplanetary discs. There is also some observational support for this. Supersonic turbulence would generate an effective disc viscosity coefficient $\alpha$ greater than unity \citep{shakura73}. Recent ALMA observations of the HL TAU disc show that mm-sized dust particles are in rather thin (geometrically) discs \citep{brogan15}, implying tiny values for the viscosity parameter, e.g., $\alpha \sim 10^{-4}$ \citep{pinte16}.

Finally, we note that we worked here with slowly cooling, $\tau_c = \beta \gg 1$, discs. In particular, $\beta$ values of $\sim 10$.  This domain is usually appropriate for protoplanetary discs in the inner $R \lesssim 50 -100$~AU \citep{rafikov05}. One can ask whether supersonic turbulence may help discs to fragment beyond that region, where the disc cooling time becomes shorter than the local dynamical time. This may well be possible. While further work is needed to explore that parameter space, we suggest that it is not likely that planet formation will be enhanced by the supersonic turbulence in the rapidly cooling (isothermal) regions of the disc. The issue here is that even if low mass gas clumps could be formed in that region of the disc (with mass well below the local Jeans mass), it is expected that these clumps will grow in mass rapidly by gas accretion in the $\beta \ll 1$ regime \citep{nayakshin17b} and will become massive brown dwarfs or even low mass stellar companions \citep{stamatellos08,kratter10,ZhuEtal12}.

\section{Conclusions}

In this paper we explored the effects of supersonic turbulence on slowly cooling protoplanetary discs. We found, in contrast to some previous work, that such discs are less likely to fragment by disc self-gravity because turbulence heats the disc strongly, rendering it more stable. In fact, our simulations and simple analytical arguments suggest that supersonic turbulence is not even possible in slowly cooling discs since the gas sound speed in such discs increases to match the rate of turbulent energy dissipation. We therefore conclude that turbulence does not make protoplanetary discs more efficient in producing planetary mass objects.

\section*{Acknowledgements}
The authors acknowledge very useful discussions with Phil Armitage and with Phil Hopkins and would like to thank the reviewer for very useful comments. KR gratefully acknowledges support from STFC grant ST/M001229/1.  The research leading to these results also received funding from the European Union Seventh Framework Programme (FP7/2007-2013) under grant agreement number 313014 (ETAEARTH). SN acknowledges support by STFC grant ST/K001000/1, the ALICE High Performance Computing Facility at the University of Leicester, and the STFC DiRAC HPC Facility (grant ST/H00856X/1 and ST/K000373/1). DiRAC is part of the National E-Infrastructure.

\label{lastpage}

\end{document}